# Synthesis of Murunskite Single Crystals: A Bridge Between Cuprates and Pnictides

Davor Tolj, Trpimir Ivšić, Ivica Živković, Konstantin Semeniuk, Edoardo Martino, Ana Akrap, Priyanka Reddy, B. Klebel-Knobloch, Ivor Lončarić, László Forró (laszlo.forro@epfl.ch)*, Neven Barišić (nbarisic@phy.hr)*, Henrik Ronnow (henrik.ronnow@epfl.ch)*, Denis K. Sunko (dks@phy.hr)*
*corresponding author

Davor Tolj, Dr. Ivica Živković, Prof. Henrik Ronnow
Laboratory for Quantum Magnetism, EPFL, 1015 Lausanne, Switzerland

Dr, Trpimir Ivšić, Dr. Konstantin Semeniuk, Dr. Edoardo Martino, Prof. László Forró
Laboratory of Physics of Complex Matter, EPFL, 1015 Lausanne, Switzerland

Prof. Ana Akrap
Department of Physics, University of Fribourg, 1700 Fribourg, Switzerland

Priyanka Reddy, Prof. Neven Barišić, Prof. Denis Karl Sunko
Department of Physics, Faculty of Science, University of Zagreb, 10000 Zagreb, Croatia

Benjamin Klebel-Knobloch, Prof. Neven Barišić
Institute of Solid State Physics, TU Wien, 1040 Vienna, Austria

Dr. Ivor Lončarić
Department of Theoretical Physics, Ruđer Bošković Institute, 10000 Zagreb, Croatia

**Abstract**

Numerous contemporary investigations in condensed matter physics are devoted to high temperature (high-$T_c$) cuprate superconductors. Despite its unique effulgence among research subjects, the enigma of the high-$T_c$ mechanism still persists. One way to advance its understanding is to discover and study new analogous systems. Here we begin a novel exploration of the natural mineral murunskite, $K_2FeCu_3S_4$, as an interpolation compound between cuprates and ferropnictides, the only known high-$T_c$ superconductors at ambient pressure. Because in-depth studies can be carried out only on single crystals, we have mastered the synthesis and growth of high quality specimens. Similar to the cuprate parent compounds, these show semiconducting behavior in resistivity and optical transmittance, and an antiferromagnetic ordering at 100 K. Spectroscopy (XPS) and calculations (DFT) concur that the sulfur 3p orbitals are partially open, making them accessible for charge manipulation, which is a prerequisite for superconductivity in analogous layered structures. DFT indicates that the valence band is more cuprate-like, while the conduction band is more pnictide-like. With appropriate doping strategies, this parent compound promises exciting future developments.

**1. Introduction**

Modern functional materials are most commonly based on ionic insulators with moderately complex chemical structure. Their physical functionality is tuned by doping, pressure, or temperature, typically without inducing structural transitions. While the functional properties do not affect binding to zeroth order, the opening of the relevant orbitals puts such materials between the textbook ionic, covalent, and metallic classes. This freedom from the burden of binding the material allows the functionalized orbitals to exhibit complex electronic behavior, with fascinating magnetic, metallic, and superconducting properties.

The prime examples of such materials are the high-temperature superconductors, cuprate perovskites and ferropnictides [1,2]. They functionalize by metallization against an ionic background. In the cuprates, the metallization is based on hole-doped oxygen orbitals [3–6]. In the pnictides, it is due to the Fe $t_{2g}$ orbitals. The $e_{2g}$ orbitals are responsible for binding to the ligands, typically arsenic, but also phosphorus [7]. While superconductivity is observed in both material families, it involves the oxygen ligands in the cuprates, but not the arsenic ligands in the pnictides [8]. Elucidating the connection between the macroscopic functional properties and the microscopic orbitals responsible for them is a major contemporary research frontier at the interface of physics and chemistry [9–14].

Along these lines, we will now focus our attention on sulfosalts. They are a mineral group which is analogous to oxides, with oxygen replaced by sulfur [15]. Compared to oxygen, sulfur is larger, less electronegative, and has d orbitals of accessible energy. Because of sulfur's smaller electronegativity, bonding to transition metals is more covalent than in oxides. In particular, this bonding can reduce the metal by creation of holes in the sulfur valence ($3p^6$) band [16]. Because sulfur has a wide range of valencies, sulfosalts have a much greater chemical variety than oxides [15]. This leads to the interesting question: can one use sulfur ligands to interpolate chemically between the cuprates and the pnictides, enabling independent manipulation of the metal and ligand orbitals?

Among sulfosalts, we choose murunskite $K_2(Fe,Cu)_4S_4$ as the material which most embodies the above idea. As **Figure 1** shows, it is structurally analogous to the much-studied ferropnictide $KFe_2As_2$ (FeAs) [17], only with As replaced by S, with the transition-metal position occupied by Fe and Cu in the ratio 1:3 in the naturally occurring mineral. This opens the possibility for two doping axes, one by changing the Fe:Cu ratio, the other by doping the potassium layer. The zeroth-order issue from this point of view is whether the sulfur orbitals in murunskite are amenable to functionalization, like oxygen in the cuprates, or they are electronically passive, like arsenic in the pnictides. The principal obstacle in the investigation of this and related questions has been the lack of any, in particular high-quality, single crystals.

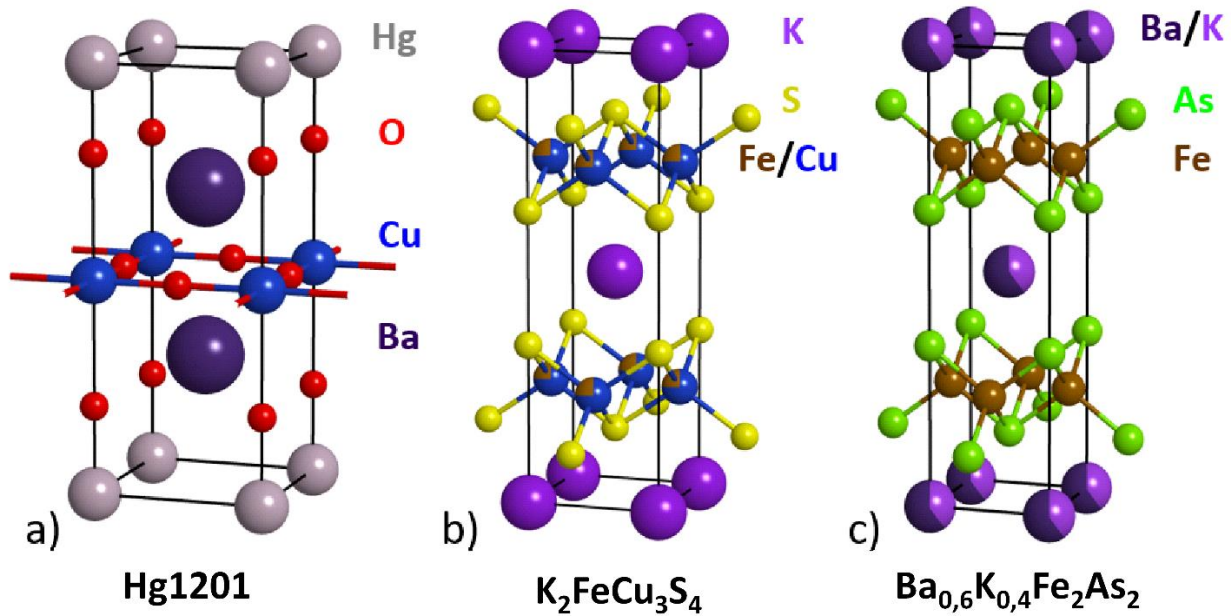

**Figure 1.** Comparison of murunskite structure b) with a cuprate a) and pnictide c). Transition metals: Cu (blue), Fe (brown). Spacers: Ba (dark purple), Hg (gray), K (purple). Ligands: O (red), S (yellow), As (green). The Fe atoms in murunskite are in the Cu positions.

Here we report on the first-ever synthesis of large (>20 mm³, >200 mg) high-quality single crystals of murunskite in the natural stoichiometry, using a new reaction pathway. They enable the application of several complementary experimental probes, e.g., Laue diffraction, X-ray photoelectron spectroscopy (XPS), high-resolution transmission electron microscopy (HR-TEM), magnetization, transport, and optical absorption. We corroborate our understanding of the data with density functional theory (DFT) and conclude that the S orbitals in murunskite are partially open even in the insulating parent compound. This implies the possibility of their functionalization, for which DFT calculations indicate a range of possible phenomena.

## 2. Synthesis and structure

Mineral murunskite is found in crystal form, typically 10 – 100 μm in size, fused in submillimeter aggregates. The original reports already noted an instability to oxidation [18]. The crystal structure was definitively established only in 2009, by diffraction on a 0.2x0.2x0.02 mm natural sample which had intercalations of water in the potassium layer [19]. A more detailed analysis was precluded by low sample quality [20]. A variant with 1:1 Fe:Cu content has been synthesized [21,22]. More recently, a hydrothermal synthesis of powders was reported, including an analogue with S replaced by Se [23].

The present single-crystal synthesis consists of two steps: in the first step, an iron copper sulfide precursor with proper ratio is prepared by solid state synthesis. Murunskite single crystals are grown in the subsequent step, directly from the melt (see experimental section).

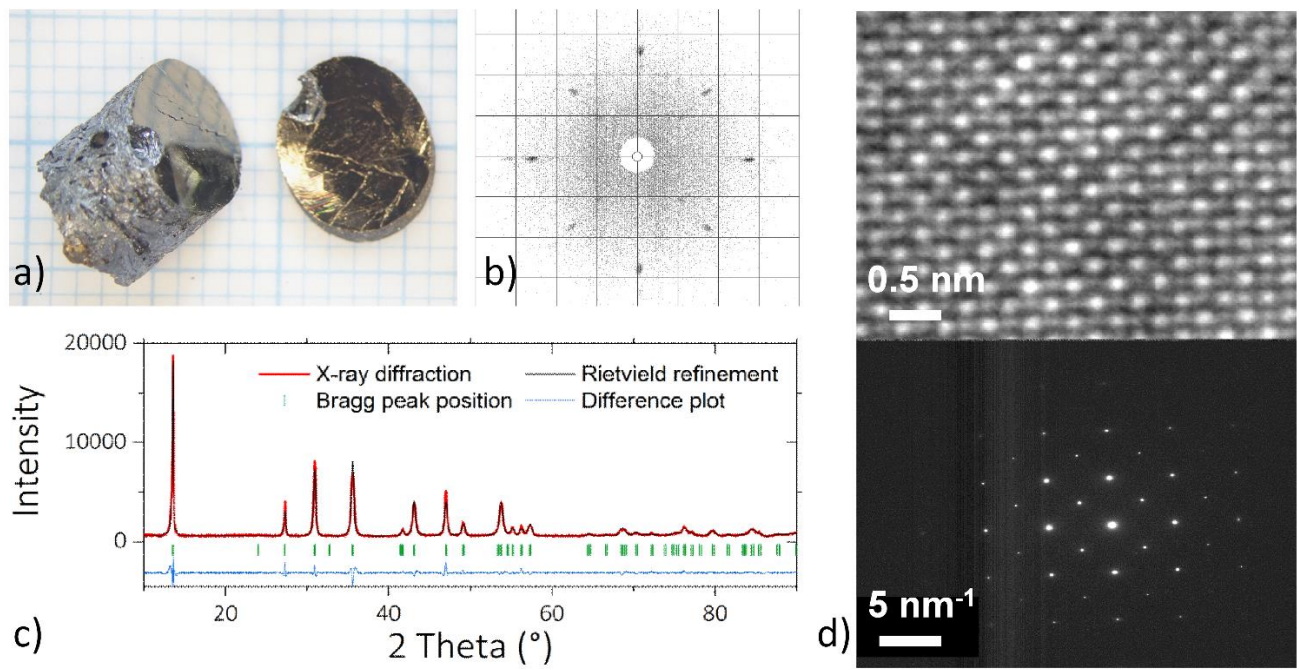

**Figure 2.** a) Optical photograph of $K_2FeCu_3S_4$ single crystals. b) Laue diffraction in c direction showing the single crystalline character of the sample. c) PXRD pattern with the difference plot between the measured and expected spectra. d) Real- and inverse-space HR-TEM images of the crystal.

Single crystals of murunskite were mechanically extracted from the bulk. Typically, they were well-formed plates with flat shiny surfaces, while the largest was a rod of dimension 7x6x10 mm$^3$ (**Figure 2a**). The c-axis Laue pattern of a cleaved plate clearly demonstrates that it is a single crystal. In real space, the HR-TEM image in Figure 2d shows a highly ordered atomic structure, corroborated by the TEM inverse-space image. In the powder x-ray diffraction pattern (PXRD), the only detected phase was the tetragonal of ThCr$_2$Si$_2$ type (space group I4/mmm, no. 139). The Rietveld refinement (R= 2.9) and the difference plot in Figure 2c, show very good agreement with the theoretical structure and no impurity phases. The lattice parameters are a = b = 3.868(1) Å, c = 13.079(9) Å, in a good agreement with reference [23]. The structure is layered and cleaves easily along the ab-plane, with potassium atoms ionically bonded to sulfur in (Cu,Fe)S$_2$ anti-PbO-type layers formed by edge-sharing MS$_4$ tetrahedra, with a random distribution of Cu and Fe atoms at the M centers. In addition to the tetrahedral sulfur shells, each M atom lies in the middle of a square of M congeners. [CCDC ###### contains the supplementary single crystal data for this paper. These data can be obtained free of charge from The Cambridge Crystallographic Data Centre via www.ccdc.cam.ac.uk/data_request/cif.]

The compositional EDX analysis of the murunskite singe crystals showed an average element ratio of K : Fe : Cu : S = 1.99 : 1.05 : 2.93 : 4.03. The composition was consistent over several samples, measured on different positions varying within the error of measurement (1 at. %).

The tendency for oxidation noted in the natural mineral depends on the sample quality. The best samples, stoichiometrically and crystallographically, are quite stable in air, showing surface morphological changes only after days of exposure. Intrusions of minority phases and departures from stoichiometry are accompanied by an accelerated formation of a black oxide layer, in agreement with the original observations [18].

## 3. Electronic properties

The results of the XPS measurements are shown in **Figure 3**, indicating the oxidation state of the constitutive atoms. In the XPS spectra for copper, the calculated full width at half maximum (FWHM) of 1.6 eV for the (2p$_{3/2}$) and 2.2 eV for the (2p$_{1/2}$) lines, along with positions of the peaks, are in good agreement with the known values for CuFeS$_2$ and KCuFeS$_2$. This confirms that only Cu$^+$ is present in murunskite, as in other reported tetragonal sulfides [22,24].

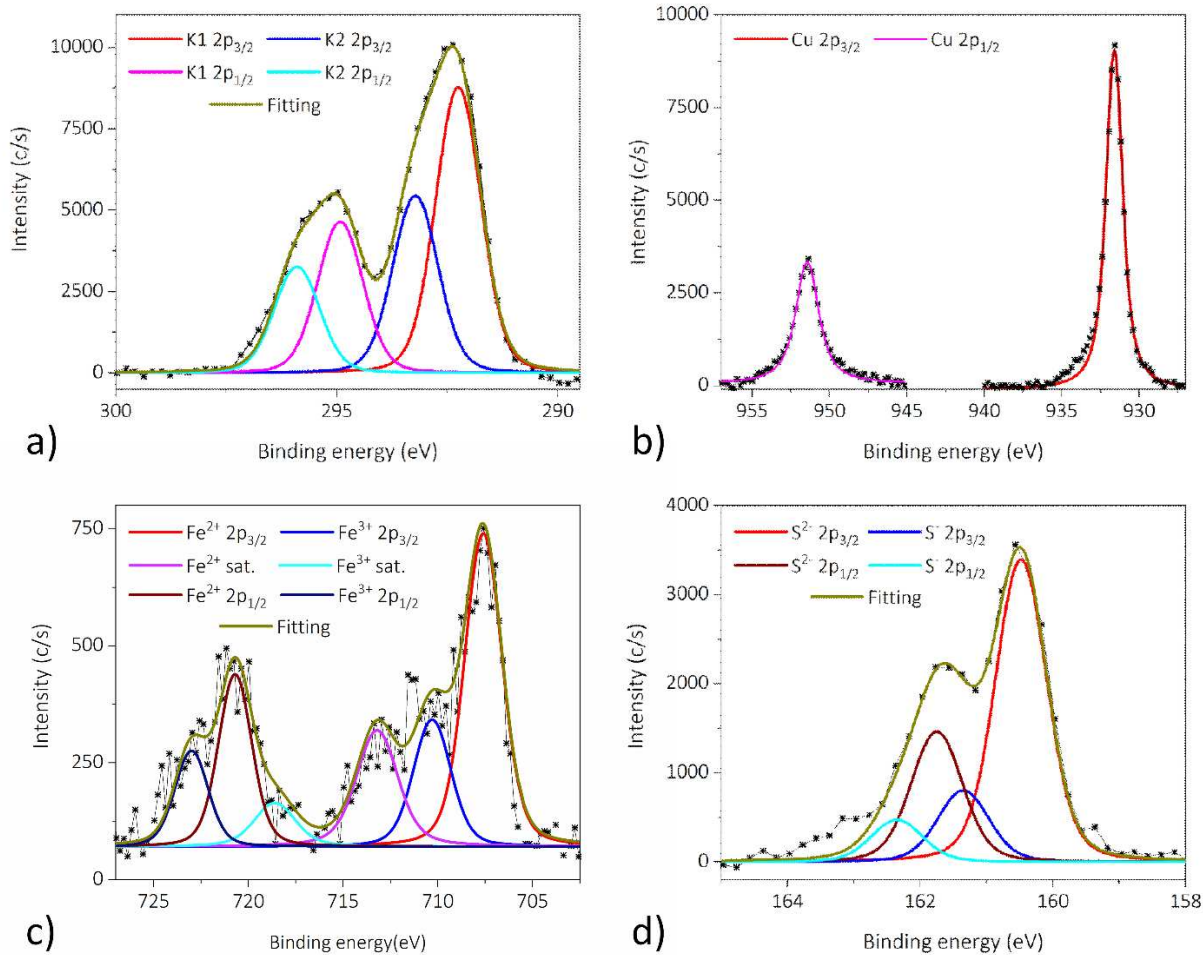

**Figure 3.** The measured and fitted XPS spectra for murunskite: a) for potassium b) copper c) iron and d) sulfur. In a) and d), two orbital states were assumed, corresponding to those observed in iron c). The respective peaks account for the observed intensity profile.

On the other hand, the spectra with binding energy values corresponding to iron $Fe^{2+}$ ($3p_{3/2}$) and $Fe^{3+}$ ($3p_{1/2}$) levels show complex features which must be fitted with multiple peaks. Simplified main and satellite peaks are shown in Figure 3c. The full fit with additional mutiplet peaks can be found in the Supporting Information (SI). The observed positions and ratio (2:1) of $3p_{3/2}$ to $3p_{1/2}$ level peaks with the existence of corresponding satellite peaks are in good agreement with the reported data indicating that iron in murunskite appears in the mixed valence state $Fe^{3+}/Fe^{2+}$ [25]. The detected $Fe^{2+}$ state is evidence that the neighboring ligand orbitals are not in the nominal oxidation state $3p^6$ either. The $Fe^{2+}$ state should be accompanied by the S $3p^5$ configuration. Direct observation of a mixed-valence state of S by XPS is complicated by the expected peak shifts being smaller than the widths. This is shown in Figure 3d, where the two extra peaks merge into an observed asymmetry of the line shapes. The respective signal is necessarily much weaker because most of the S atoms are in the vicinity of $Cu^+$, whose ionicity is nominal. A similar asymmetry of the line shape is also observed in potassium XPS spectra. Experimental data for K $3p_{3/2}$ and $3p_{1/2}$ level peaks must be fitted by two sets of peaks (K1, K2) as shown in Figure 3a. Because the potassium is ionically coordinated by sulfur, two different potassium environments are inferred, in accordance with XPS data for sulfur and iron. Hence, we conservatively conclude that XPS sees a single oxidation state $S^{2-}$ at present, corresponding to the $3p^6$ configuration, while allowing the presence of the state $S^-$ to be inferred from the line shapes. Further evidence of a $3p^5$ ($S^-$) component is provided by magnetic measurements, to which we now turn.

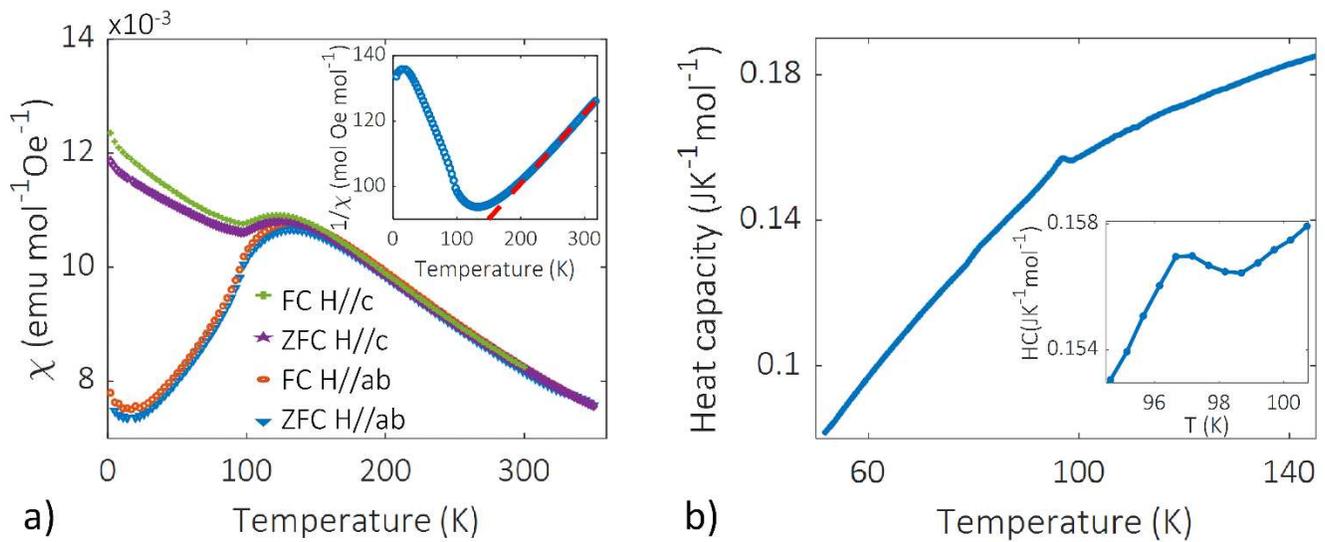

**Figure 4.** a) Temperature dependence of the magnetic susceptibility χ with H = 1 T applied (FC) and zero field (ZFC) runs. The inset shows the fitted result using the Curie–Weiss law, where the red line is the fitting curve for FC, H ∥ ab measurement. b) Heat capacity versus temperature, with a pronounced peak around the AFM transition, magnified in the inset.

Magnetic susceptibility temperature dependence is presented in **Figure 4**a. Paramagnetic behavior at high temperatures shows a Curie-Weiss-like behavior. It is followed by a broad maximum characteristic for the appearance of strong magnetic correlations. No difference was observed between field cooled (FC) and zero field cooled (ZFC) protocols, suggesting an antiferromagnetic structure. Magnetization measurements also display strong anisotropy. There is a significant difference between in-plane and out-of-plane magnetic behavior in Figure 4a, suggesting that magnetic moments are oriented along the ab plane. The temperature dependent magnetic susceptibilities were fitted to a Curie-Weiss law in the temperature range ≈170-350 K,

$$\chi(T) = \chi_0 + \frac{C}{T-\theta}, \tag{1}$$

where $C$ is the Curie constant and $\theta$ the Curie-Weiss temperature. The magnetic moment was assumed to be on iron atoms, because alkali metals have no unpaired electrons, and copper ions are in the non-magnetic $d^{10}$ state in sulfides [24]. The best fit to the experimental data (see inset to Figure 4a) yields $\theta$ = -269 K, indicating strongly antiferromagnetic interactions and C = 4.95 emu mol$^{-1}$ Oe$^{-1}$ K$^{-1}$ corresponding to an effective magnetic moment of 6.3 $\mu_B$. Calculations assuming spin-only contribution give 5.92 and 4.89 $\mu_B$ for $Fe^{3+}$ and $Fe^{2+}$. Experimentally observed magnetic moments in other transition metal compounds with spin 5/2 usually deviate from calculated moments and range between 5.7 $\mu_B$ and 6.1 $\mu_B$. The observed 6.3 $\mu_B$ in murunskite is therefore at this range's upper end. The moment's large magnitude can additionally be attributed to a strong covalent mixing of the sulfur 3p and iron 3d bands near the Fermi level, leading to increased electron density around iron, and decreased at sulfur sites. This mixing effectively amounts to additional unpaired spins in the $FeCu_3S_4$ sulfide layer, therefore increasing the observed aggregate magnetic moment, and pointing further to the ligand orbital's electronic activity. Such a scenario is supported by the DFT calculations below, in which the S orbitals are partially open and carry a small magnetic moment. The phase transition at 100 K seen in susceptibility measurements is also clearly observed in the heat capacity in Figure 4b. The overall temperature dependence is that of the lattice contribution, and the peak in the inset emphasizes its jump at the phase transition.

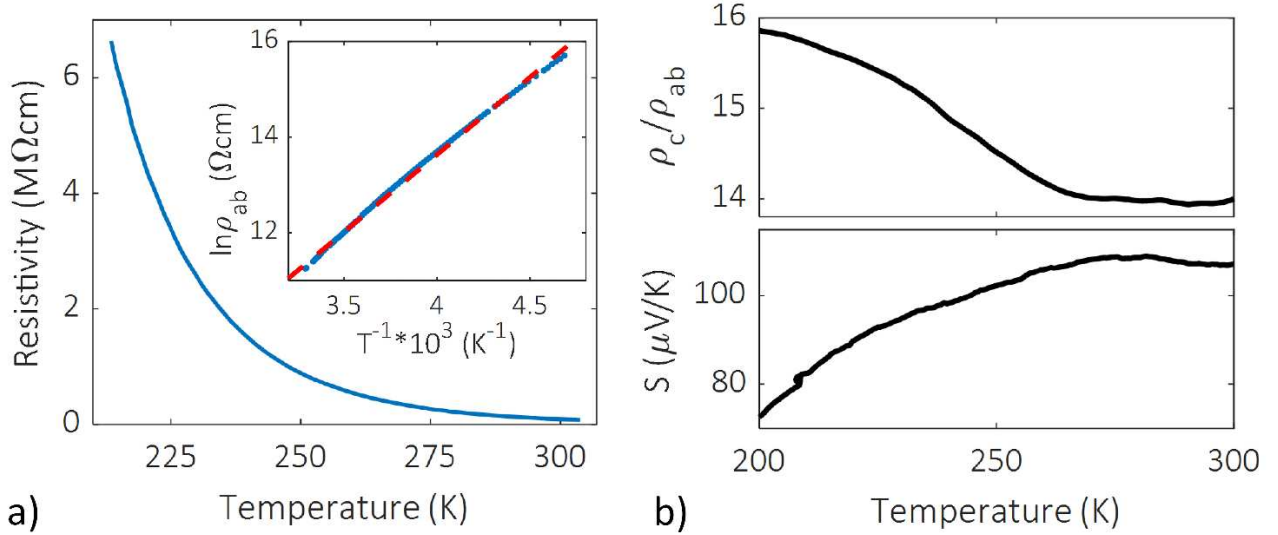

**Figure 5.** a) Temperature dependence of the in-plane resistivity of $K_2FeCu_3S_4$ single crystal. Inset: the plot $\ln\rho$ vs $T^{-1}$ used to extract the activation energy for conduction. b) (upper panel) The inter- and in-plan resistivities, $\rho_c/\rho_{ab}$ in the 200-300 K range. b) (lower panel) The temperature dependence of the Seebeck coefficient, showing predominantly hole conduction.

**Figure 5**a presents the temperature dependence of the in-plane electrical resistivity ($\rho_{ab}$) for a $K_2FeCu_3S_4$ single crystal. It shows semiconducting behavior in the accessible temperature range 210 to 300 K, with a room-temperature value of 0.6 MΩcm. Below 210 K, the resistance was too large to be measured, testifying incidentally to the low impurity content of the crystal. The thermal activation model was fitted using the equation

$$\rho_{ab} = \rho_0 \exp\left(\frac{E_a}{k_B T}\right) \quad (2)$$

where $\rho_0$ is the prefactor, $k_B$ is the Boltzmann constant and $E_a$ is the thermal activation energy, fitted to 0.3 eV. The observed activation energy would be half the gap value in an intrinsic semiconductor. This simple inference of 0.6 eV for the intrinsic gap is subject to some caution. The position of the chemical potential is affected by the open ligand orbitals, which may lead to an asymmetry between particle and hole conduction even in the insulating state. In such a 'self-doping' scenario, the material behaves more like an extrinsically doped semiconductor, with the chemical potential closer to one edge of the gap. In this case, the thermal activation energy is the smaller of the two distances from the chemical potential to the edges of the gap.

It is possible that the intrinsic disorder in Cu and Fe positions plays a role in charge transport, judging from the straight line of the log $\rho_{ab}$ versus $T^{-1/4}$ plot (see SI). In the high-temperature context, it should be interpreted as multi-phonon assisted hopping [26].

In Figure 5b (upper panel), the ratio between the inter- and in-plane resistivities, $\rho_c/\rho_{ab}$ is plotted. Its room temperature value is 15, which increases upon cooling. Due to the layered structure the conduction anisotropy is even observed in this pristine compound, in which holes are the dominant type of charge carriers, as shown by the Seebeck coefficient (S) in Figure 5b (lower panel). The temperature dependence characteristic of a semiconductor (S ~ 1/T) is not observed, pointing to the possible role of electrons as minority carriers.

To further elucidate the relationship between the activation energy and gap size, **Figure 6** compares the optical transmittance (TR) measured at ambient pressure with the optical absorption calculated in DFT. The measured TR extrapolates to zero near 0.8 eV, indicating that the value of 0.6 eV inferred from activated conduction was an underestimate. Further evidence in the same direction comes from DFT calculations.

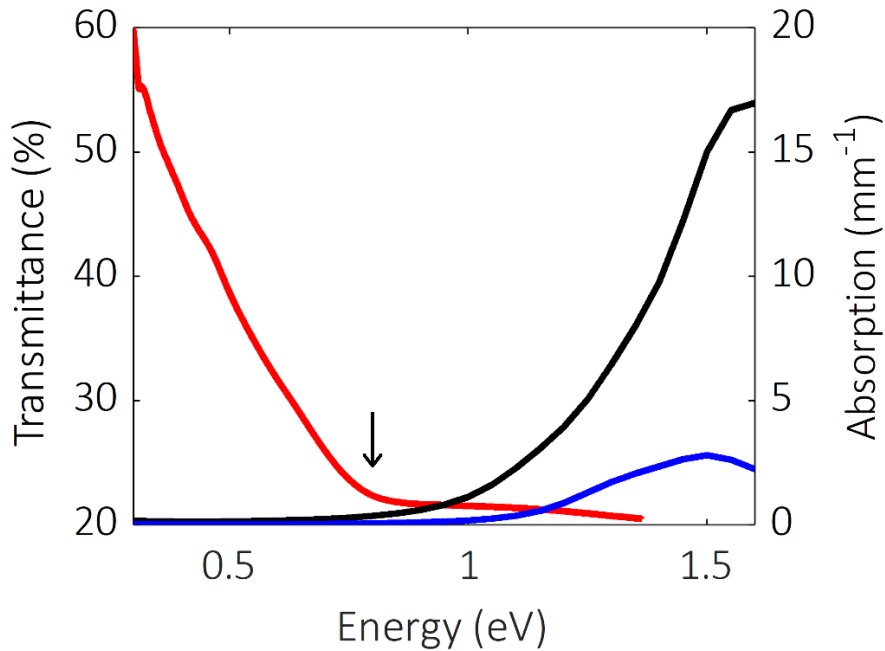

**Figure 6.** Measured transmittance at ambient pressure (red line) compared to the absorption coefficient as calculated from the xx and zz components of the dielectric tensor in DFT. The optical gap, defined by the extrapolated transmittance, is at 0.8 eV, marked with an arrow.

We have calculated the band structure of the pristine compound in DFT with QuantumATK code, using the Medium basis set and PBE exchange-correlation functional [27–29]. The Brillouin zone was sampled with a 13x13x8 Monkhorst-Pack grid. To reproduce the 3:1 Cu:Fe ratio, the ab-plane unit cell was quadrupled to 2x2. We then unfolded the bands into the large zone which does not distinguish between the Cu and Fe atoms in crystallographically equivalent positions, as in the lattice depicted in Figure 1. Because Fe and Cu are randomly distributed in the real material, such a projection, originally developed for disordered alloys, corresponds to what would be observed in ARPES, up to matrix-element effects [30]. The band structure is given in **Figure 7**a. This calculation a priori predicts an insulating band gap of 0.4 eV along with a $Fe^{3+}$ ($3d^5$) configuration, in contradiction with the experiment. We added a standard +U correction for intra-orbital Coulomb repulsion, $U_d = 6$ eV within the Fe d orbitals, which pushed a hole to the S ligand, reducing the iron to $Fe^{2+}$ in accord with XPS observations, and widening the gap to 0.95 eV in the density of states (DOS), given in Figure 7b. A slight broadening of single-particle peaks, which stands in for Fe/Cu disorder effects in the above-mentioned band unfolding, brought the gap down to 0.86 eV in the simulated ARPES spectra, in very good agreement with the 0.8 eV inferred from optical transmittance.

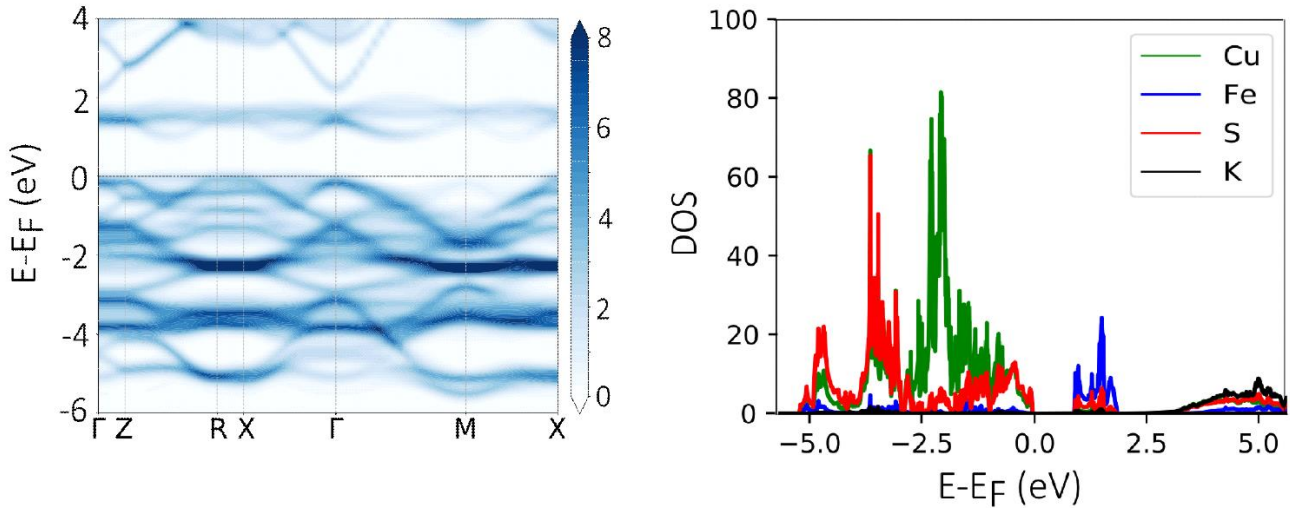

**Figure 7.** DFT results for a) the band structure and b) DOS. In a) the unfolded band structure is given with an intensity map of the bands projected onto the observable zone, as explained in the text. The notable feature in b) is the large S content immediately below the gap, hybridized with Cu. Bands above the gap are dominated by Fe. The gap in the DOS is 0.95 eV. The indirect gap in the band structure a) is narrower (0.86 eV) because of level broadening in the projection. See the text for a discussion of the chemical potential.

The main qualitative insight from the calculation is that ligand orbitals significantly contribute to the lowest-energy bands immediately below the gap. It is also interesting to observe that the valence and conduction bands are copper and iron-dominated, respectively. The calculation indicates that the corresponding hole and electron metals would be quite different, one making the material more analogous to cuprates, the other to pnictides.

Because the +U correction of 6 eV in the DFT calculation was adjusted to produce the observed iron 2+ state, the reasonable DFT gap estimate is an independent prediction of the calculation. The relatively slight underestimate of its value by activated conduction is probably not significant. However, it is in accord with the possibility that the chemical potential is not strictly in the middle of the gap either, along the moderate self-doping scenario discussed above. In this context, it is notable that in cuprates, both hole and electron doping induce hole conductivity.[6,31]

| Atomic orbital configuration | Calculated filling: total (up–down) | Nominal ionic configuration |
|:---:|:---:|:---:|
| K 3p6 4s1 | 3p: 5.98 (0.00) 4s: 0.19 (0.01) | 3p6 4s0 |
| Cu 3d10 4s1 | 3d: 9.81 (0.00) 4s: 0.56 (0.00) | 3d10 4s0 |
| Fe 3d6 4s2 | 3d: 6.17 (4.20) 4s: 0.60 (0.10) | 3d5 4s0 |
| S 3s2 3p4 | 3s: 1.78 (0.00) 3p: 4.77 (0.14) | 3s2 3p6 |

**Table 1.** Comparison of calculated (Mulliken) and nominal ionicities for the open atomic orbitals, with orbital polarizations. Fe is calculated to be 2+ vs. the nominal 3+, with S in the corresponding 2p5 configuration. The +U correction is 6 eV on Fe d orbitals alone.

The calculated ionicities are given in **Table 1**. Atomic charges were estimated by standard Mulliken population analysis [32]. The Fe d orbitals depart significantly from the nominal ionicities, indicating a 2+ state. The charges allow us to infer the spins on the atoms from general considerations. One expects a high-spin state for Fe in the tetrahedral coordination, which amounts to S=2 for the calculated filling $3d^6$, which is the dominant oxidation state $Fe^{2+}$ observed in XPS. The calculation indicates that the observed larger magnetic moment is related to a partial polarization of the S orbitals, predicting that they are electronically active even in the insulating parent compound.

## 4. Discussion

The present work views sulfur ligand orbitals as a potential pathway for functionalization in materials based on murunskite as a parent compound. Whether these orbitals are accessible to chemical and physical manipulation at all is the fundamental question we addressed. Magnetization and XPS investigations, supported by DFT calculations, concur that the sulfur orbitals are partially open in the parent insulator, and are significantly present in the lowest-energy bands immediately below the optical gap. The resolution of this critical issue is positive.

Evidence that the S orbitals are partially open hinges on a concurrence between XPS measurements and DFT calculations. XPS finds a mixed-valence state of Fe, with a preponderance of $Fe^{2+}$ over $Fe^{3+}$, the exact ratio not relevant for this discussion. Given that XPS is a surface probe, we cannot strictly exclude the possibility that the bulk material is of a single orbital state. However, if one is interested in doping, the main point is simply that the fully ionic $3d^5$ $Fe^{3+}$ state is easily suppressed in favor of the $Fe^{2+}$ state with $3d^6$ configuration. The latter is reproduced by our DFT calculation with a reasonable +U correction on the Fe d orbitals. In the calculation, Fe is reduced by electron transfer from S, which ends in a $3p^5$ state. Notably, the calculated DOS shows a large S content in the valence (hole) bands below the gap. Surface oxidation, if present, would push Fe towards the 3+ state, as indeed observed in XPS of visibly oxidized samples. Thus, we conclude that the observed $Fe^{2+}$ state is on account of opening the S $3p^6$ configuration, no matter why the reduction occurs.

Having said that, it is also reasonable to believe that the XPS measurements reflect a mixed-orbital state of the bulk material, because Fe and Cu are randomly distributed despite their different nominal ionicities. The simplest way to account for this randomness is that the Fe atoms draw electrons from S orbitals at no cost to compensate for the local Cu/Fe environment, in accord with the ligands being more covalent than ionic. The same self-doping effect could account for the gap inferred from activated conduction being slightly smaller than the one from optical transmission and DFT calculation, because it allows for the chemical potential to be not exactly in the middle of the insulating gap. Such an interpretation is supported by the valence and conduction bands being Cu- and Fe-dominated, respectively, in the DFT calculation, naturally allowing for a corresponding asymmetry between hole and electron conduction.

Magnetization measurements indicate an excess of magnetic moment in the bulk, over that of Fe alone. The difference could arise from polarized S 3p orbitals. The DFT calculation also gives a polarized S $3p^5$ orbital along with the $Fe^{2+}$ state. Thus, all available evidence is in favor of a bulk mixed-valence $Fe^{2+/3+}$ state accompanied with partially open S orbitals, even if the quantitative polarization estimates are subject to change. The detection of hole-like charge carriers by the Seebeck coefficient supports this notion.

Murunskite is structurally analogous to the ferropnictide $KFe_2As_2$, in which the ligand As orbitals are deep below the Fermi level, and the parent compound is metallic via the Fe $t_{2g}$ orbitals, which do not hybridize with the ligands at all. On the other hand, in the octahedrally coordinated cuprates, the ligand O orbitals metallize immediately upon doping [33,34], and are significantly present in the lowest-energy bands. We find that the structural analogy does not hold up microscopically, because the valence band of murunskite is electronically more similar to cuprates than pnictides. On the other hand, the conduction (empty) band is pnictide-like, literally realizing the idea of murunskite as an interpolation compound. The difference to cuprates is that the partially open S orbitals are still insulating in the parent compound. This indicates that a competition between ionicity and covalency is a unifying narrative for functional materials.

## 5. Conclusions and Perspectives

We have elaborated a synthesis route for large single crystals of murunskite by a new reaction pathway. This allows the in-depth study of the compounds electronic and magnetic properties, opening the possibility to the material doping. Magnetization, optical, and XPS investigations, supported by DFT calculations, indicate that the ligand S orbitals are partially open in the parent insulating compound, appearing along with a mixed-valence $Fe^{2+/3+}$ orbital state. Despite the structural similarity to ferropnictides, murunskite is electronically more closely related to cuprates, opening interesting possibilities for the functionalization of ligand orbitals. This indicates additional possibilities for doping, which we plan to explore in the next phase of our investigations.

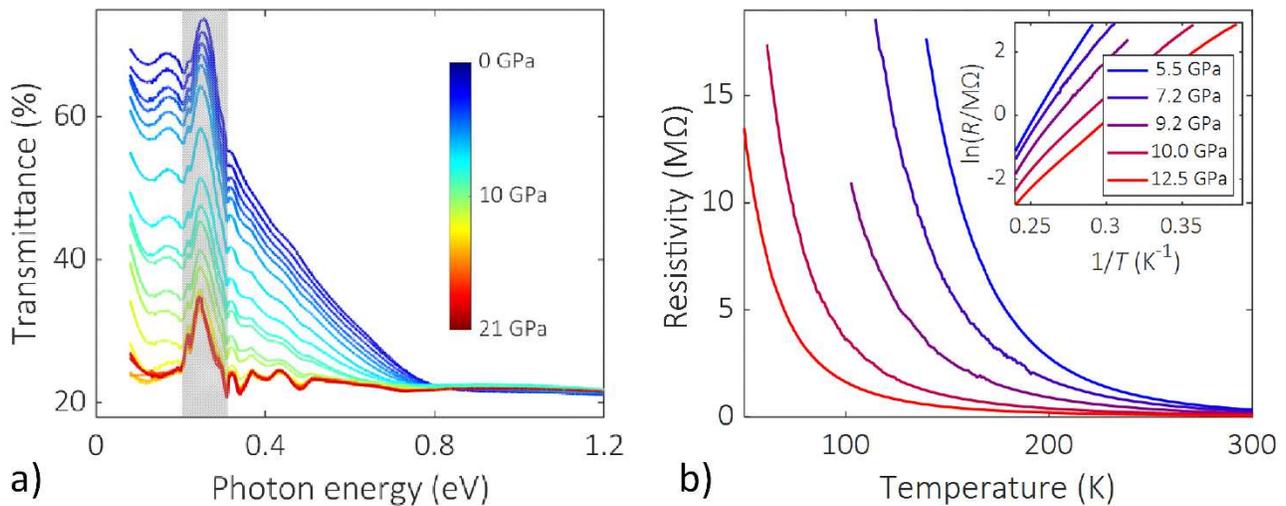

**Figure 8.** Pressure dependence of a) the optical transmission and b) the d.c. resistivity in a DAC. The inset gives the Arrhenius plot of resistivity. Both data sets show a rapid increase of the conductivity of the material with pressure. (The greyed-out curves from 0.2 to 0.3 eV show absorption by the diamond anvils, and may be ignored.)

One particularly attractive possibility is self-doping, which could be explored through high pressure studies. The mechanism is that increasing the Coulomb interactions by compressing the lattice can redistribute charges in the ligand orbitals, making the compound more conducting, and ultimately superconducting. Such an effect is more likely in layered materials, where the compressibility in the c-direction is high. Test studies of optical transmission and d.c. resistivity (**Figure 8**) show an unusually strong attenuation of the charge transfer gap with pressure. The strong decrease of the transmittance corroborates a progressive closing of the charge transfer gap. A similar behavior is observed in the resistivity (Figure 8b). The open question at present is which transition-metal orbitals will metallize first, those hybridized with ligands, or others. Superconductivity is known to occur by the first of these pathways in the cuprates, and by the second in the pnictides. To resolve this question, higher pressures would likely be needed. The rapid decrease of optical transmittance below 0.8 eV without significant change in the extrapolated energy at which it becomes strictly zero (the optical gap by definition) is reminiscent of a similar "filling without closing" of the SC gap, observed in cuprates [35].

In perspective, much is expected from versatile chemical doping, by selectively creating charge carriers in the valence (hole-like) and conduction (electron-like) bands. On the route to superconductivity by these doping strategies, exciting novel phenomena are expected in murunskite.

**Acknowledgements**

Technical help of Bi Wen Hua (David) is gratefully acknowledged. The work at the University of Zagreb was supported by project CeNIKS co-financed by the Croatian Government and the European Union through the European Regional Development Fund -Competitiveness and Cohesion Operational Program (Grant No. KK.01.1.1.02.0013), the Croatian-Swiss Research Program of the Croatian Science Foundation and the Swiss National Science Foundation with funds obtained from the Swiss-Croatian Cooperation Program Project No. IZHRZ0_180652, and by the Croatian Science Foundation under Project No. IP-2018-01-7828. The work at TU Wien was supported by the European Research Council (ERC Consolidator Grant No. 725521). A. A. acknowledges funding from the Swiss National Science Foundation through project PP00P2 170544. A part of this work was done at the SMIS beamline of Synchrotron Soleil, Proposal 20181880.

**Experimental section**

Synthesis

Iron copper sulfide precursor was prepared by solid state synthesis. Powders of elemental iron, copper and sulfur (purity at least 99.5%) in molar ratios 1:3:4, respectively, were mixed in 5 g batches, pressed into rods by a hydrostatic press, sealed in evacuated quartz tubes and heated at 450 °C for 24 h. After cooling down, the batch was ground in an argon atmosphere, pressed into rods again, sealed in evacuated quartz tubes and annealed at 700 °C for 48 h. The obtained bulk copper iron sulfide was reground into powder in an argon atmosphere and directly used in the next step.

Single crystals of murunskite were grown from the melt. Iron copper sulfide precursor powder (1g) was mixed in an agate mortar with stoichiometric amount of pure potassium metal (purity 99.9%). The powder was then sealed in evacuated quartz tubes, heated and kept at 930 °C for 12 h. Afterwards the melt was slowly cooled to 800 °C at the rate of 2 °C h$^{-1}$, after which heating was stopped and the product was left to cool down to room temperature.

X-ray study

Precursors, powder and single crystals of murunskite were characterized by powder X-ray diffraction (PXRD) at room temperature on a diffractometer with Cu K$\alpha$ radiation ($\lambda$ = 1.5148 Å) operating at 45 kV and 0 mA and a diffracted-beam graphite monochromator in a reflection mode (2$\theta$ = 10–90°, step = 0.013° 2$\theta$). Small amounts of precursors and murunskite single crystals samples were thoroughly powderized in an argon glovebox and placed on a low background holder for measurement.

A crystal of murunskite of suitable size was selected for single crystal x-ray diffraction (SCXRD) and mounted on a duo-source X-ray diffractometer (source used was Mo K$\alpha$, $\lambda$ = 0.71073 Å). After determining the unit cell, full data collection was performed at 100K to obtain 99% of completeness with resolution up to 0.65 Angstrom. After solving the structure, it was refined by minimizing the full-matrix least-squares against F-square. The occupancy factors were fixed according to the results of EDX measurements, with all atoms in the asymmetric unit cell constrained to have the same atomic displacement parameters.

[CCDC ###### contains the supplementary crystallographic data for this paper. These data can be obtained free of charge from The Cambridge Crystallographic Data Centre via www.ccdc.cam.ac.uk/data_request/cif.]

Compositional analysis

Sample morphology and composition information were observed by scanning electron microscopy (SEM). Energy-dispersive X-ray spectroscopy (EDX) was performed on multiple single crystal samples to get precise elemental ratio information.

X-Ray Photoelectron Spectroscopy (XPS) measurements were analyzed using a monochromatic Al Kα X-ray source of 24.8 W power with a beam size of 100 µm. The spherical capacitor analyzer was set at 45° take-off angle with respect to the sample surface.

Magnetic measurements

Magnetization as functions of temperature and applied magnetic field was measured on single crystal using a MPMS-5T superconducting quantum interference device magnetometer (SQUID).

Electrical measurements

A conventional four-point technique was employed to measure the resistivity. Indium wires were mechanically pressed on the sample to minimize the Schottky barrier at the interface (contacts made with silver paste exhibited very high contact resistance).

High-pressure resistivity measurements were conducted in a diamond anvil cell, using a four-point technique. NaCl powder was used as a pressure transmitting medium. Measurement leads were directly pressed into the sample, without using any conducting adhesives. Pressure was recorded at the end of each temperature sweep, and was determined from the shifts of the R1 fluorescence line of ruby, pieces of which were placed into the sample space.

Optical transmission

High pressure Infrared transmittance measurements were performed on the infrared beamline SMIS at the synchrotron SOLEIL. Spectra collection were performed on a custom-built horizontal microscope for diamond anvil cells (DACs), equipped with custom Cassegrain objectives, a MCT and a Si diode detectors, for the Mid-IR and the Near-IR respectively [36]. The horizontal microscope was coupled to Thermo Fisher iS50 interferometer with Quartz (NIR) / KBr (MIR) beam splitters and synchrotron radiation as IR source.

Square samples of about 100 × 100 µm2 were loaded into a membrane Diamond anvil cell with 400 µm culets. Stainless-steel gaskets were pre-indented to a thickness of about 50 µm and a hole of diameter 150 um was drilled by electro-erosion. NaCl was used as pressure transmitting medium [37]. Pressure was measured in situ by the standard ruby fluorescence technique. Data was collected every 1 GPa during both compression and decompression, with the last data point collected after the cell was open to release any residual pressure.

**Conflict of Interest**

The authors declare no conflict of interest.

**Supporting Information and Data Availability**

Supporting information and data are available from the corresponding authors on reasonable demand.

# Supporting Information

**Synthesis of Murunskite Single Crystals: A Bridge Between Cuprates and Pnictides**

Davor Tolj, Trpimir Ivšić, Ivica Živković, Konstantin Semeniuk, Edoardo Martino, Ana Akrap, Priyanka Reddy, B. Klebel-Knobloch, Ivor Lončarić, László Forró*, Neven Barišić*, Henrik Ronnow*, Denis K. Sunko*

| Formula | $K_{1.95}Fe_{1.10}Cu_{2.85}S_{4.15}$ |
|---|---|
| Formula weight | 450.28 |
| Crystal size (mm$^3$) | 0.10x0.08x0.04 |
| Space group | *I4/mmm* (no. 139) |
| $a$ (Å) | 3.8477(2) |
| $c$ (Å) | 13.0213(12) |
| $V$ (Å$^3$) | 192.778(3) |
| Z | 1 |
| $P_{calc}$ (g cm$^{-3}$) | 3.879 |
| $\lambda$ (Mo K$\alpha$) (Å) | 0.71073 |
| Temperature (K) | 100 |
| $M_u$ | 11.844 |
| $2\theta_{max}$ (deg) | 66 |
| Scan type | $\omega$-$2\theta$ |
| Unique reflections | 137 |
| F000 | 214.0 |
| F000' | 216.13 |
| $(h,k,l)_{max}$ | 5,5,20 |
| Goodness of fit, S | 1.038 |
| Rp, Rwp | 0.0190, 0.0427 |

**Table S1** Crystallographic data and experimental details for $K_{1.95}Fe_{1.10}Cu_{2.85}S_{4.15}$

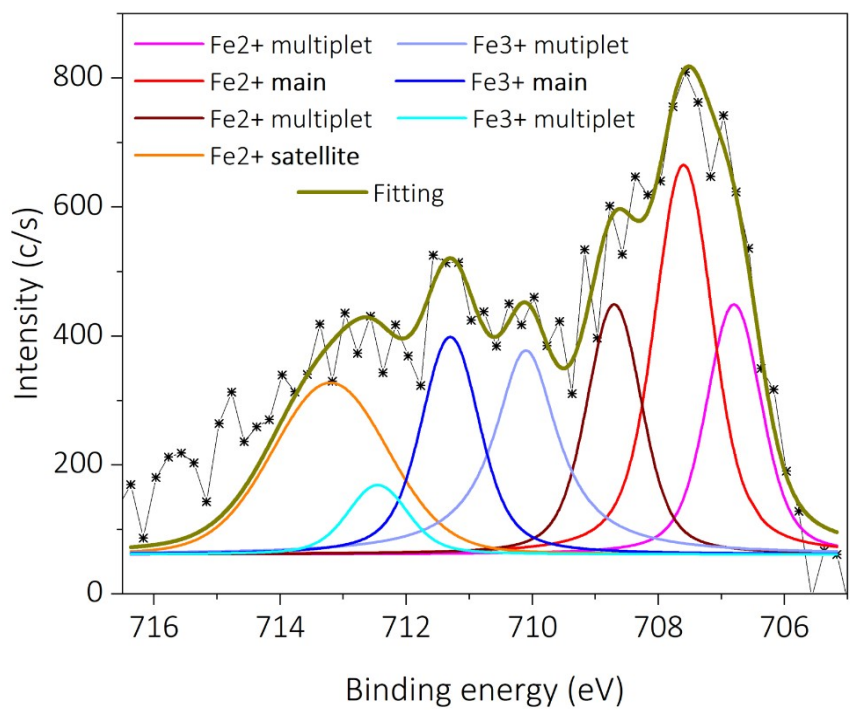

**Figure S1** The measured and full fit for iron XPS spectra with multiplet peaks included

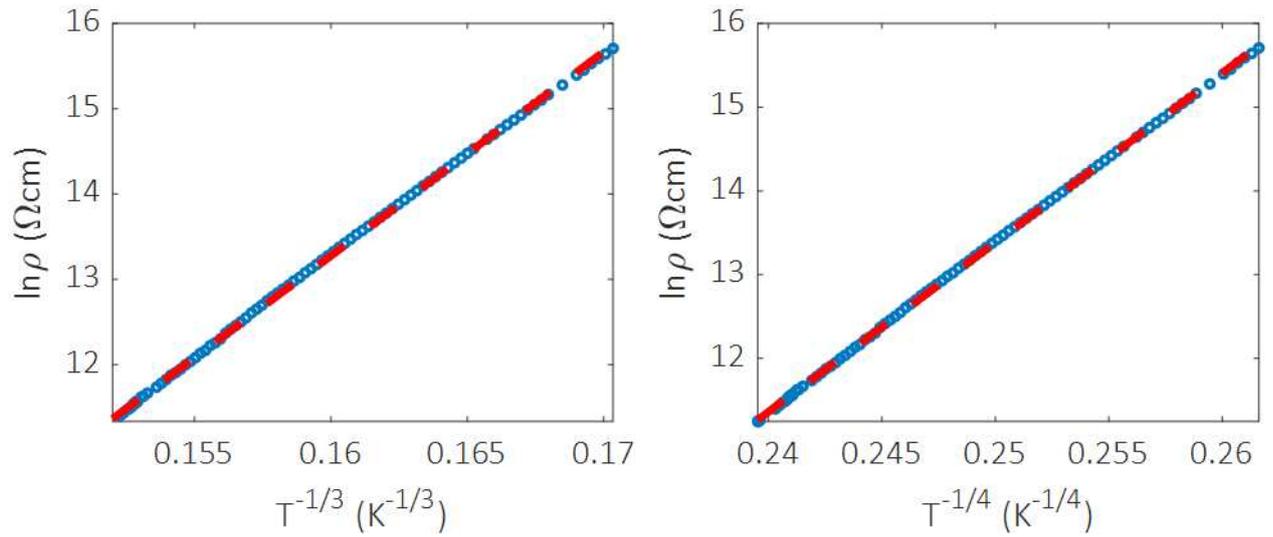

**Figure S2** The plot a) ln$\rho_{ab}$ vs T$^{-1/3}$ and b) ln$\rho_{ab}$ vs T$^{-1/4}$ showing good fit to experimental data (with T$^{-1/4}$ having slightly better goodness of fit)

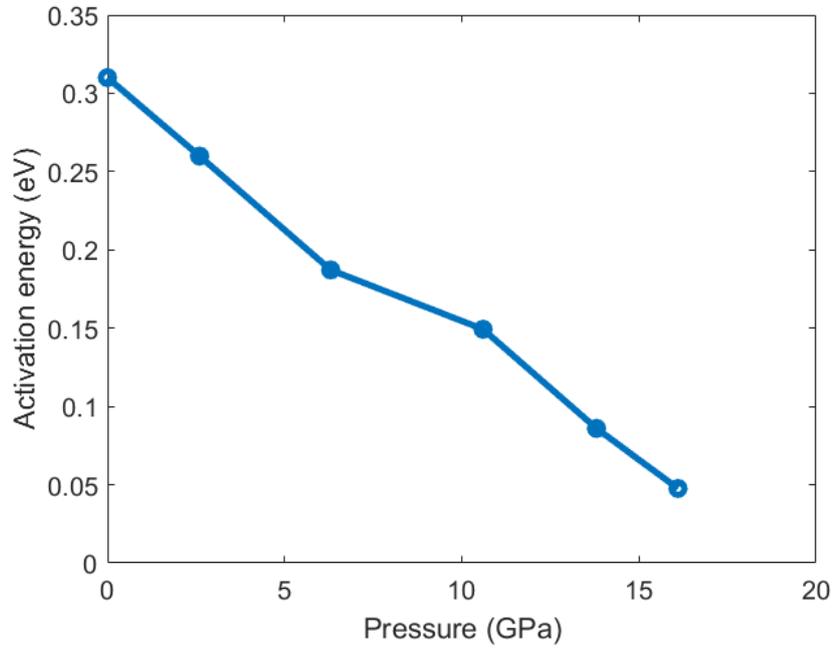

**Figure S3** Pressure dependance of Activation energy (Ea) observed from resistivity measurements calculated by thermal activation model